\documentclass[prb,aps,twocolumn,showpacs,nobibnotes,epsf]{revtex4}

\usepackage{mathptmx}
\usepackage{bm}
\usepackage{graphicx,amssymb,amsmath}

\begin{document}


\title{A crossover in phase diagram of NaFe$_{1-x}$Co$_x$As determined by electronic transport measurements }

\author{A. F. Wang}
\author{J. J. Ying}
\author{X. G. Luo}
\author{Y. J. Yan}
\affiliation{Hefei National Laboratory for Physical Science at
Microscale and Department of Physics,
University of Science and Technology of China, Hefei, Anhui 230026,
People's Republic of China}

\author{D. Y. Liu}
\affiliation{Key Laboratory of Materials Physics, Institute
of Solid State Physics, Chinese Academy of Sciences, P. O. Box 1129,
Hefei 230031, People's Republic of
China}

\author{Z. J. Xiang}
\author{P. Cheng}
\author{G. J. Ye}
\affiliation{Hefei National Laboratory for Physical Science at
Microscale and Department of Physics,
University of Science and Technology of China, Hefei, Anhui 230026,
People's Republic of China}

\author{L. J. Zou}
\affiliation{Key Laboratory of Materials Physics, Institute
of Solid State Physics, Chinese Academy of Sciences, P. O. Box 1129,
Hefei 230031, People's Republic of
China}

\author{Z. Sun}
\affiliation{National Synchrotron Radiation Laboratory, University of Science and Technology of China, Hefei, Anhui 230029, People's Republic of China}

\author{X. H. Chen}
\altaffiliation{Corresponding author} \email{chenxh@ustc.edu.cn}
\affiliation{Hefei National Laboratory for Physical Science at
Microscale and Department of Physics, University of Science and
Technology of China, Hefei, Anhui 230026, People's Republic of
China}

\begin{abstract}
We report electronic transport measurements on single crystals of
NaFe$_{1-x}$Co$_x$As system.  We found that the cotangent of Hall
angle, cot$\theta_{\rm H}$, follows $T^4$ for the parent compound
with filamentary superconductivity and $T^2$ for the
heavily-overdoped non-superconducting sample. While it exhibits
approximately $T^3$-dependence in all the superconducting samples,
suggesting this behavior is associated with bulk superconductivity
in ferropnictides. A deviation develops below a characteristic
temperature $T^*$ well above the structural and superconducting
transitions, accompanied by a departure from power-law temperature
dependence in resistivity. The doping dependence of $T^*$ resembles
the crossover line of pseudogap phase in cuprates.
\end{abstract}


\maketitle

\vskip 25 pt
\section{INTRODUCTION}
The ferropnictide and cuprate high-temperature superconductors share
some key similarities - an antiferromagentism in parent compounds,
the quasi-two-dimensional nature of superconducting CuO$_2$ and FeAs
layers \cite{kamihara,chen,cruz,sefat}, and the emerging
superconductivity realized by suppressing the antiferromagnetic
ground states \cite{lee,greene}. A natural question is whether the
two families have the same mechanism for superconductivity. In
cuprate compounds, the nature of pseudogap phase is a key issue in
understanding the high-temperature superconductivity. The evidence
from resistivity \cite{ando}, Nernst effect \cite{daou}, inelastic
neutron scattering \cite{stock,hinkov1,hinkov2} and scanning
tunnelling spectroscopy \cite{kohsaka1,kohsaka2} indicated that the
pseudogap phase in cuprates is an electronic state that breaks
rotational symmetry of underlying lattice with stripe or nematic
order. Compared with the phase diagram of cuprates, the current
interest in ferropnictides lies in the peculiar normal-state
properties of these materials to detect whether there exists a
pseudogap-like state.

In ferropnictides,  the electronic nematicity has been detected by
in-plane resistivity anisotropy \cite{chu}, angle-resolved photoemission measurements  \cite{yi},  scanning tunnelling spectroscopy
\cite{chuang} and inelastic neutron scattering \cite{zhao}. In these
studies, the structural distortions that break the crystal's $C_4$
rotational symmetry may apply external driving forces to induce the
electronic nematicity. Very recently, magnetic torque measurements
by Kasahara \textit{et al.} revealed a nematic transition at high
temperatures above the structural and superconducting transitions in
BaFe$_2$(As$_{1-x}$P$_x$)$_2$, with a characteristic temperature similar
to the pseudogap crossover in cuprate superconductors
\cite{kasahara}.  Without the potential external driving forces from
the lattice, this transition can be considered to be induced by the
electronic system only.

However, there lacks evidence directly related to the electronic
properties  of the high-temperature nematic transition reported by
Kasahara \textit{et al.}. For an evident phase transition, such as
the superconducting, antiferromagnetic and structural transitions in
ferropnictides, specific heat data show anomalies at corresponding
temperatures.  It becomes nontrivial to locate the crossover line
and determine the phase diagram. In this letter, we study the
resistivity and Hall coefficient measurements on high-quality single
crystals of NaFe$_{1-x}$Co$_x$As. At high temperatures, the Hall
angle, $cot \theta_{\rm H}$, reveals $T^3$-dependence for
superconducting samples. Below a crossover temperature $T^*$, the
resistivity and Hall angle deviate from power-law temperature
dependence. In the electronic phase diagram, $T^*$ depicts a
characteristic temperature similar to the pseudogap crossover in
cuprate superconductors.

\begin{figure}
\centering
\includegraphics[width=0.50\textwidth]{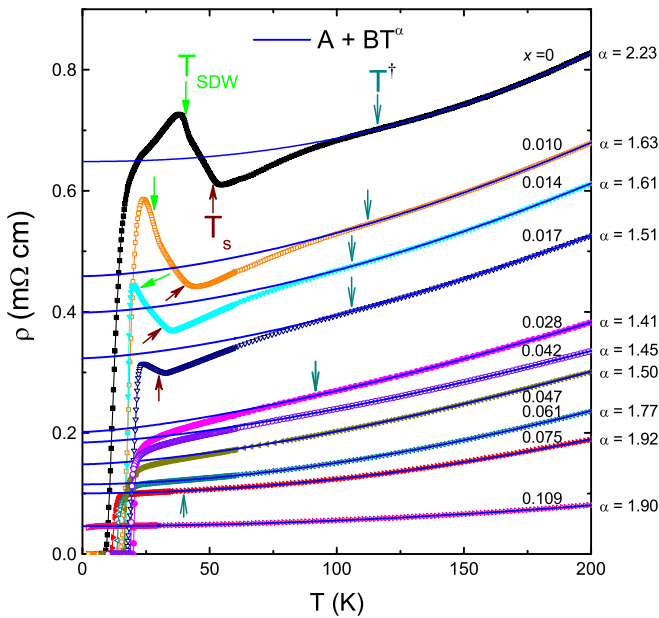}
\caption{Temperature dependence of in-plane resistivity for
NaFe$_{1-x}$Co$_x$As single crystals. The solid line is the fitting
curve by power-law temperature dependence with the formula:
$\rho=A+B$$\times$$T^\alpha$ to the resistivity data. The power-law
exponents, $\alpha$, are shown on the right of the panel, which
decrease from 2.23 for $x$ = 0 to 1.41 for the optimal doping level
and then increase to 1.90 for the heavily overdoped sample. Obvious
deviation from the high-temperature power-law behavior can be seen
in resistivity and the deviation temperature ($T^{\dagger}$)
decreases with increasing Co concentration. In addition, the
characteristic temperatures of SDW and structural transitions are
also marked as $T_{SDW}$ and $T_s$ }
\end{figure}

\begin{figure*}
\centering
\includegraphics[width=\textwidth]{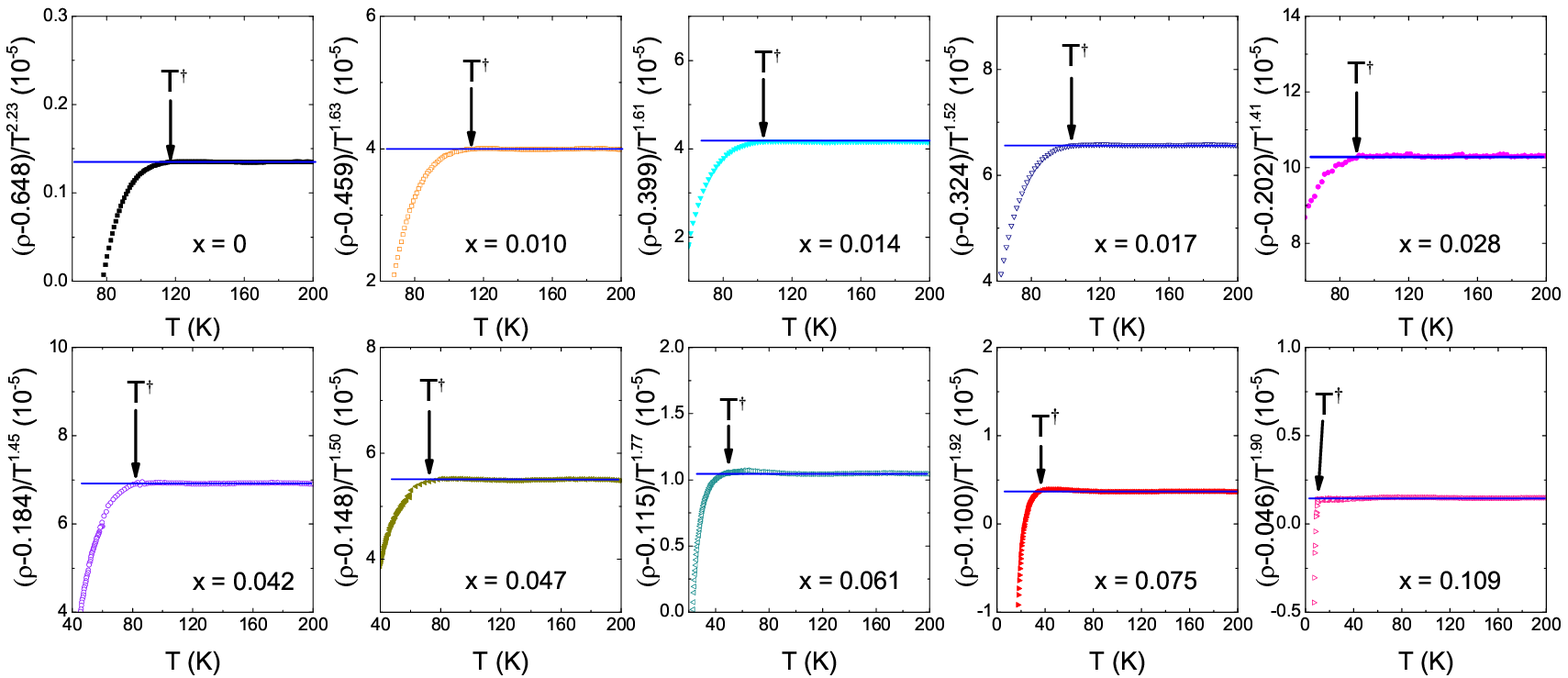}
\caption{($\rho - A$)/$T^{\alpha}$ as a function of temperature for
all the crystals of  NaFe$_{1-x}$Co$_x$As. It clearly shows the
temperature at which the deviation from the high-temperature
power-law dependence of in-plane resistivity $\rho$ . The dashed
lines guide eyes for the the $A$+$B$$\times$$T^\alpha$ dependence.
The deviation at $T^\dag$ is marked by arrows.}
\end{figure*}

\begin{figure}
\includegraphics[width=0.45\textwidth]{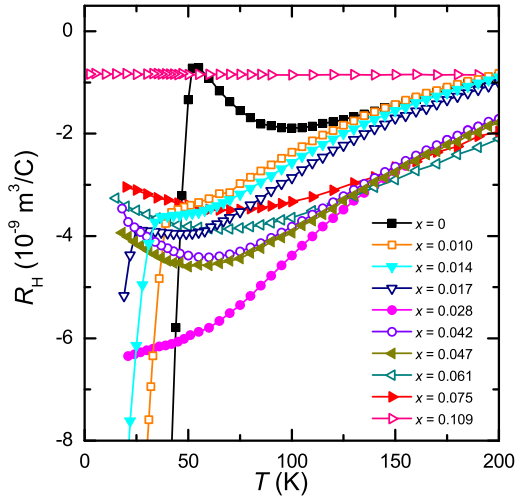}
\caption{Temperature dependence of Hall coefficient, $R_H$, for
various single crystals of NaFe$_{1-x}$Co$_x$As.}
\end{figure}

\begin{figure*}
\centering
\includegraphics[width=0.9\textwidth]{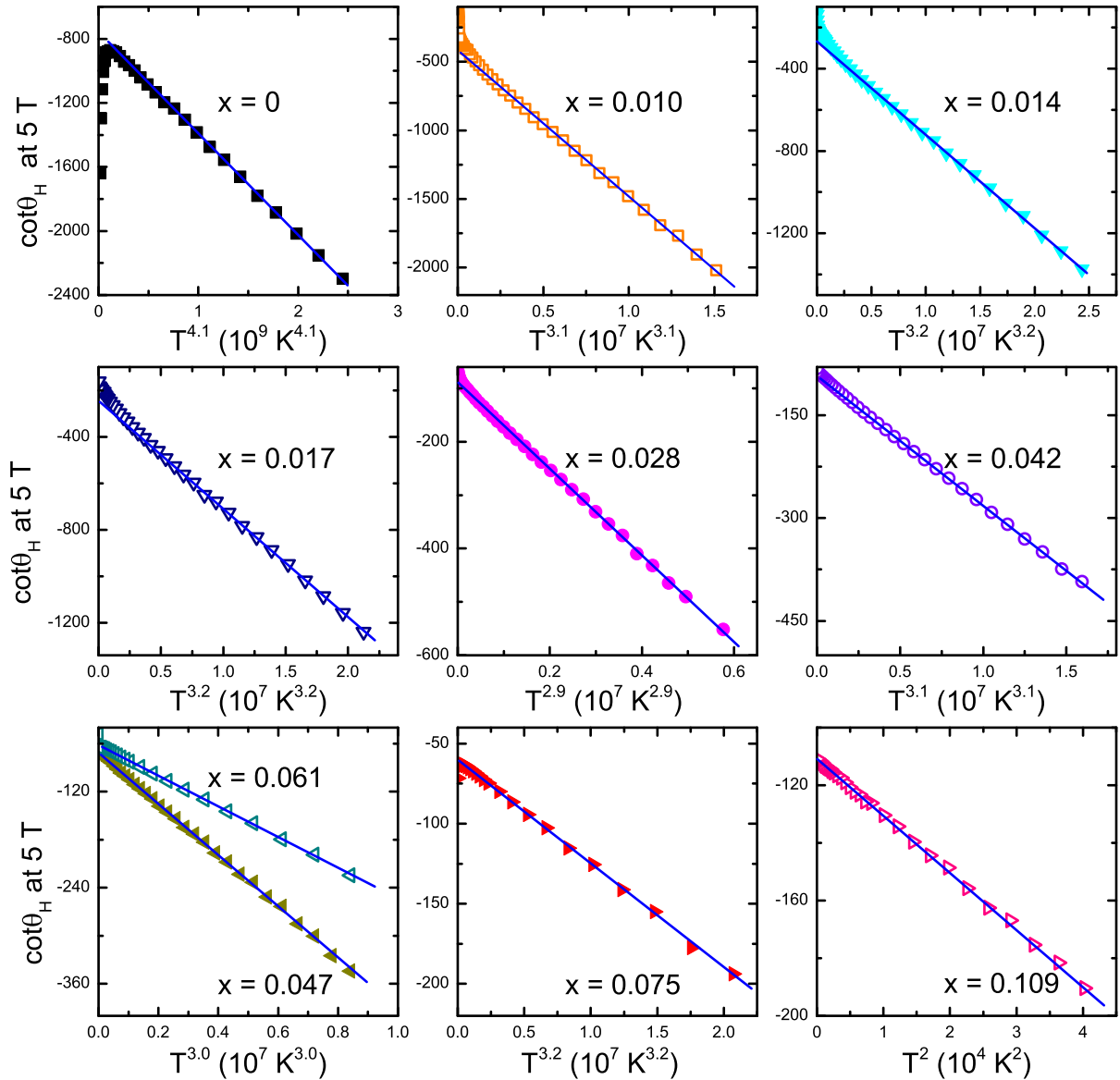}
\caption{The cot$\theta_{\rm H}$ plotted in power-law temperature
scale for NaFe$_{1-x}$Co$_x$As single crystals. All the
superconducting samples show approximately $T^3$ dependent
cot$\theta_{\rm H}$ except for the parent compound with filamentary
superconductivity. }
\end{figure*}

\begin{figure*}
\centering
\includegraphics[width=\textwidth]{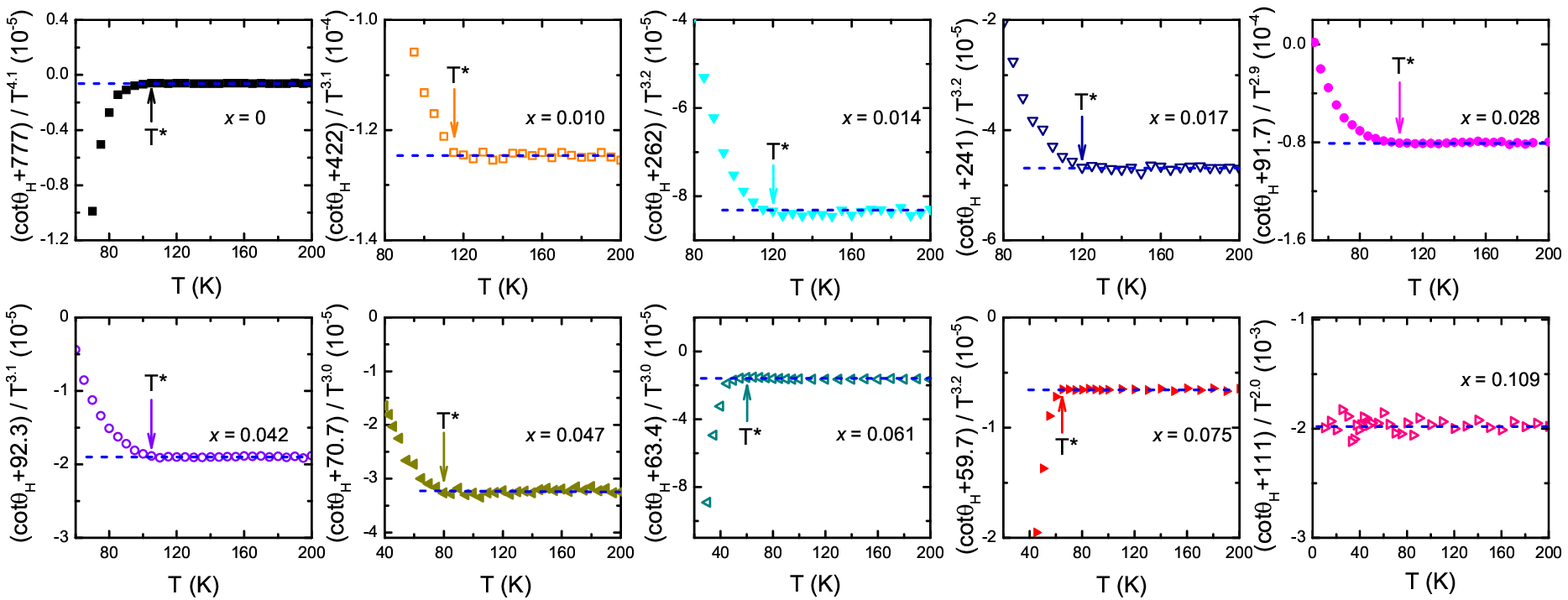}
\caption{(cot$\theta_{\rm H}$ - $C$)/$T^{\beta}$ as a function of
temperature for all the crystals of  NaFe$_{1-x}$Co$_x$As. It
clearly shows the temperature at which the deviation from the
high-temperature power-law dependent cot$\theta_{\rm H}$. The dashed
lines guide eyes for the the $C$+$D$$\times$$T^\beta$ dependence.
The deviation at $T^\ast$ is marked by arrows.}
\end{figure*}

\begin{figure}[ht]
\includegraphics[width=0.45\textwidth]{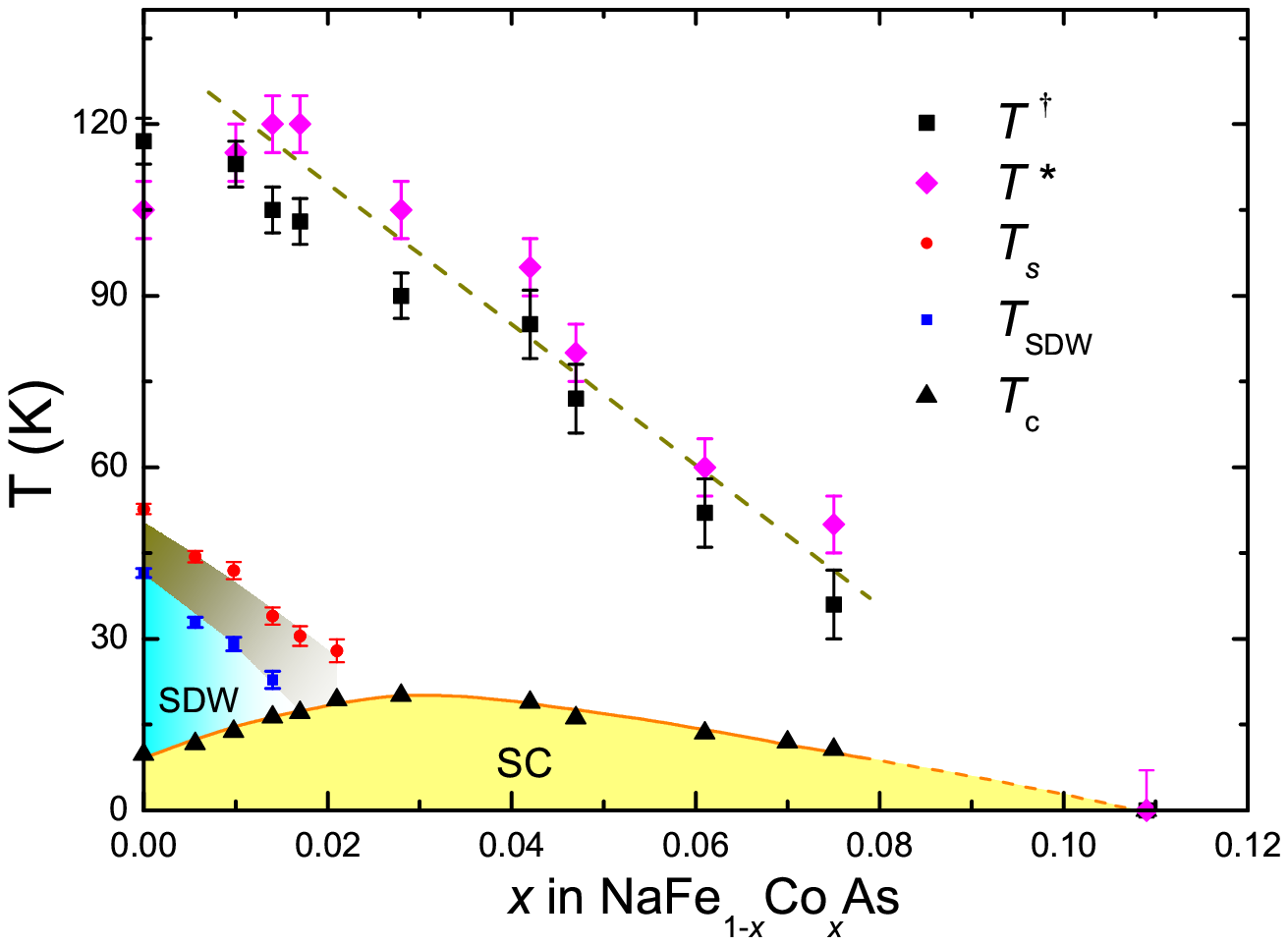}
\caption{Phase diagram of NaFe$_{1-x}$Co$_x$As system. The
superconducting transition temperature, $T_c$, spin density wave
(SDW) transition temperature, $T_{SDW}$, and structural transition
temperature, $T_s$, were determined from resistivity, susceptibility
and specific heat measurements by our group \cite{waf}. For the
parent compound, the specific heat shows the structural and SDW
transitions with no anomaly at $T_c$, suggesting a filamentary
superconductivity. $T^{\dag}$ and $T^{\ast}$ were determined in
Figs. 2 and Fig. 5, respectively.}
\end{figure}
\section{EXPERIMENTAL DETAILS}
High-quality single crystals of NaFe$_{1-x}$Co$_x$As were grown by
the NaAs flux method \cite{waf}. The accurate chemical composition
of the single crystals was determined by energy dispersive X-ray
spectroscopy (EDS). The standard instrument error for this method is
around 10\%. The single crystals of $x=0$, 0.010, 0.014, 0.017,
0.028, 0.042, 0.047, 0.061, 0.075 and 0.109 were measured, with a
high quality characterized and demonstrated in our previous report
\cite{waf}. X-ray diffraction (XRD) was performed on Smartlab-9
diffractometer (Rikagu) from 10$^{\rm o}$ to 60$^{\rm o}$ with a
scanning rate of 2$^{\rm o}$/minute. Measurements of resistivity and
Hall effect were conducted by using the PPMS-9T (Quantum Design).
The resistivity were measured using the standard four-probe method.
The contacts cover the sides of the samples to ensure in-plane
transport. Hall coefficient was measured by sweeping the field from
-5T to 5T at various temperatures. We firstly confirmed the
linearity of Hall resistivity $\rho_{xy}$ to magnetic field in the
normal state. An excellent linearity can be observed for all the
samples in the normal state except that a very weak deviation from
the linearity appears below $T_{\rm SDW}$ in underdoped samples.
Then the temperature dependence of Hall coefficients was obtained
from the substraction of the voltages measured at 5 T and -5 T,
$R_{\rm H}$ = [$V_{xy}$(5T)-$V_{xy}$(-5T)]$\times d$/2$I$, where $d$
is the thickness of crystals and $I$ is the current. The cotangent
of Hall angle, cot$\theta_{\rm H}$=$\rho/\rho_{xy}$, was calculated
at 5 T, where $\rho_{xy}$ is Hall resistivity. It should be
addressed that the Hall coefficient can not be well measured at the
temperature above 200 K, the possible reason is that Na ions could
move above 200 K. All the samples we used are the same with that in
our previous work\cite{waf} and the $T_{\rm c}$, $T_{\rm SDW}$ and
$T_{\rm s}$ are summarized in Table 1 and plotted in Fig. 6. All the
superconducting samples show nearly full shielding fraction except
for the parent compound with filamentary superconductivity.

\section{RESULTS AND DISCUSSION}
We carefully measured the resistivity on single crystals of
$NaFe_{1-x}Co_xAs$ system with $x=0-0.109$ ranging from the parent
compound to the heavily overdoped non-superconducting composition as
shown in Fig. 1. For the parent compound and the underdoped
crystals, there exists an upturn in resistivity below 50 K due to
the structural and spin density wave (SDW) transitions \cite{waf}.
An evident feature is that the resistivity curvature changes well
above the $T_{\rm s}$ in parent compound and all underdoped samples.
For the optimally doped and overdoped crystals, there seems no
anomaly in the normal-state resistivity. In the high-temperature
cuprate superconductors, the mapping of in-plane resistivity
curvature is a useful way to determine electronic phase diagrams. In
particular, the pseudogap crossover line can be conveniently
obtained by this method \cite{ando1}. Here, we made power law
fitting on the resistivity data with formula $\rho=A+B \times
T^\alpha$ and show them in Fig.1. One can see that the resistivity
follows power-law dependence at high temperature and starts to
deviate at a characteristic temperature, $T^{\dag}$, for all the
samples. The doping dependence of $T^{\dag}$ will be shown in Fig.
6. For the heavily overdoped non-superconducting sample ($x=0.109$),
no deviation from the power-law dependence can be observed. As shown
in Fig. 1, the power-law exponent $\alpha$ decreases with increasing
Co doping from the parent compound to the optimally doped crystals,
and then increases with Co concentration in the overdoped region.
The $\alpha$ reaches the smallest value of 1.41 in the optimally
doped sample.

To accurately determine the $T^\dag$, we plot ($\rho$-A)/$T^\alpha$
\textit{vs}. temperature, as shown in Fig. 2. The characteristic
temperature $T^\dag$ can be well defined, at which $\rho$ starts to
deviate from the power-law temperature dependence. As shown in Fig.
2, the heavily overdoped crystal with x=0.109 shows a deviation from
the power-law temperature dependence due to a tiny superconducting
transition around 6 K, so that we take the $T^\dag$=0 K. The
deviation temperature, $T^\dag$, determined by our fitting is highly
repeatable. Table 1 summarizes the $T^\dag$, which monotonically
decreases with increasing Co concentration and goes to zero in the
heavily overdoped non-superconducting compound ($x$ = 0.109). We
should addressed that the variation of the temperature region used
for fitting does not  change the deviation temperature $T^\dag$,
significantly, which suggests that our results are reliable.

Figure 3 shows the temperature dependence of Hall coefficients,
$R_{\rm H}$, for various single crystals of NaFe$_{1-x}$Co$_x$As.
The Hall coefficients show a systematic evolution with increasing Co
doping with a strong temperature dependence. The magnitude of Hall
coefficients at room temperature increases upon Co doping, and then
sharply decreases for the heavily overdoped non-superconducting
sample with a value smaller than those of superconducting samples.
It is worth noting that the Hall coefficient in the
non-superconducting $x=0.109$ compound is nearly independent of
temperature, bearing a feature that is usually found in a
conventional Fermi-liquid metal. This behavior suggests that
NaFe$_{1-x}$Co$_x$As becomes a traditional metallic material at very
high doping levels. The abrupt increase below 50 K in magnitude of
Hall coefficients for the underdoped samples is due to the
structural transition.

The complicated properties of temperature dependent Hall
coefficients can be expressed in a simple fashion by looking at the
cotangent of Hall angle: cot${\theta}_{\rm H}=\rho/\rho_{xy}$
\cite{chien} as shown in Fig. 4, where $\rho_{xy}$ is Hall
resistivity. It is interesting to notice that cot${\theta}_{\rm H}$
shows power-law temperature dependence for all the single crystals
of NaFe$_{1-x}$Co$_x$As: $T^4$ for the parent compound,
approximately $T^3$ for all the superconducting crystals, and $T^2$
for the heavily-overdoped non-superconducting sample.
$T^\beta$-dependent cot${\theta}_{\rm H}$ with $\beta$=2.5 $\sim$
3.0 has been reported in BaFe$_{2-x}$Co$_x$As$_2$ single crystals
\cite{Arushanov}. In cuprates, cot${\theta}_{\rm H}$ behaves
approximately as $T^2$ \cite{chien}, regardless of materials and
doping level, except that $T^4$-dependence is found in the
electron-doped cuprates \cite{Wang}, which is interpreted by the
multi-band effect with different contributions from various bands.

A careful examination on Fig. 4 indicates slight curvatures in the
plot, suggesting that the best power-laws in a wide temperature
range deviate slightly from an integer $\beta$. Fig. 5 shows plots
of (cot$\theta_{\rm H}-C$)/$T^{\beta}$ \textit{vs.} $T$ for all the
samples, in which the power-law temperature dependence is canceled
out, so that one can easily see the temperature range in which the
$T^\beta$ behavior holds well. Here, $C$ is the offset value and
$\beta$ is the best power. In Fig. 5, the power-law temperature
dependent cot$\theta_{\rm H}$ holds very well down to a
characteristic temperature, $T^\ast$, for all the samples. Below
$T^*$, cot$\theta_{\rm H}$ departs from the power-law ($T^\beta$)
behavior. This characteristic temperature decreases with increasing
Co doping and goes to zero in the heavily overdoped
non-superconducting sample. One may notice that the deviation goes
in different directions for various doping, which is probably
related to the details in the quasiparticle scattering processes
that varies with Co concentrations. We should addressed that the
variation of the temperature region used for fitting does not change
the deviation temperature $T^\ast$, significantly, which suggests
that our results are reliable.

\begin{table}
\centering \caption{The temperatures plotted in the phase diagram
(see Fig. 6 in the manuscript).}\label{Table 1}
\begin{tabular}{cccccc}\hline\hline
$x$   & $T_{\rm c}$ (K) & $T_{\rm SDW}$ (K) & $T_{\rm s}$ (K)  &
$T^\ast$ (K) & $T^\dag$ (K) \\ \hline
0     & 9.8$\pm$0.3     & 40.6$\pm$1.0 & 51.4$\pm$1.0 & 105$\pm$5 & 116$\pm$4 \\
0.010 & 13.8$\pm$0.3    & 28.2$\pm$1.0 & 41.0$\pm$1.5 & 115$\pm$5 & 113$\pm$4 \\
0.014 & 16.3$\pm$0.3    & 21.8$\pm$1.5 & 32.1$\pm$1.5 & 120$\pm$5 & 106$\pm$4 \\
0.017 & 17.1$\pm$0.3    &              & 30.0$\pm$1.8 & 120$\pm$5 & 103$\pm$4 \\
0.028 & 20.1$\pm$0.3    &              &              & 105$\pm$5 & 92$\pm$4 \\
0.042 & 18.9$\pm$0.3    &              &              & 95$\pm$5  & 85$\pm$6\\
0.047 & 16.2$\pm$0.3    &              &              & 80$\pm$5  & 70$\pm$6 \\
0.061 & 13.5$\pm$0.3    &              &              & 60$\pm$5  & 52$\pm$6 \\
0.075 & 10.7$\pm$0.3    &              &              & 50$\pm$5  & 40$\pm$6 \\
0.109 & 0               &              &              &  0        & 0    \\
\hline \hline
\end{tabular}
\end{table}

Fig. 6 shows the phase diagram and we plot $T^{\dag}$ and $T^\ast$
as a function of doping. The two characteristic temperatures are
highly consistent with each other, though they are obtained by
different methods. Generally, both $T^{\dag}$ and $T^\ast$ decreases
with increasing Co doping, and goes to zero at the doping level $x$
= 0.109 where $T_{\rm c}$ goes to zero. The phase diagram in Fig. 6
is quite similar to the pseudogap phase diagram of the
high-temperature cuprate superconductors. Indeed, the deviation of
cot $\theta_{\rm H}$ from $T^2$ dependence has been used to
characterize the onset of pseudogap in hole-doped cuprate
superconductors \cite{ando2}, and the pseudogap crossover line can
be conveniently determined by resistivity curvature mapping
\cite{ando1}. Here, these methods were used to determine the
$T^{\dag}$ and $T^\ast$ and reveal the crossover line above the
structural and superconducting transitions. Our data strongly
suggests that a crossover occurs at the $T^\ast$ in ferropnictide
superconductors. We are cautious that our data do not indicate any
"gapping" behavior as observed in cuprates, though we stress the
similarity between characteristic $T^\ast$ temperature and the
pseudogap crossover temperature

A natural question is what happens below the $T^\ast$. Without the
driving forces from structural or magnetic transitions, the $T^\ast$
indicates a crossover with purely electronic origin. A possible
scenario is that the electronic nematic state sets in below the
characteristic temperature. Indeed, a similar crossover induced by
electronic nematicity has been observed in BaFe$_2
$(As$_{1-x}$P$_x$)$_2$ well above the structural and superconducting
transitions\cite{kasahara}. The Hall angle exhibits approximate
$T^3$-dependence for all the superconducting crystals, while the
parent compound with filamentary superconductivity and the heavily
overdoped non-superconducting crystal exhibit a different behavior.
This finding suggests that the $T^3$-dependence of cot$\theta_{\rm
H}$ is tied to the superconducting region.

\section{CONCLUSION}
In conclusion, Our electronic transport studies reveal the existence
of a crossover temperature well above structural and magnetic
transitions. In the phase diagram of $NaFe_{1-x}Co_xAs$, the
crossover line resembles the pseudogap phase diagram in cuprate
superconductors. An interesting phenomenon is that the Hall angle
reveals approximate $T^3$-dependence of cot$\theta_{\rm H}$ in the
whole superconducting regime, suggesting this behaivor is associated
with bulk superconductivity in ferropnictides. These findings shed
light on the mechanism of the high-temperature superconductivity in
ferropnictides and potentially the superconductivity in cuprates as
well.

{\bf ACKNOWLEDGEMENTS}
X.H. Chen would like to thank Yayu Wang,
Zheng-Yu Weng and J. P. Hu for stimulating discussion and
suggestion. This work is supported by National Natural Science
Foundation of China (Grant No. 11190021, 51021091), the National
Basic Research Program of China (973 Program, Grant No. 2012CB922002
and No. 2011CBA00101), and Chinese Academy of Sciences.

\end{document}